# A data-driven approach to spoken dialog segmentation

David Griol[a], José Manuel Molina[a], Araceli Sanchis[a], Zoraida Callejas[b]

[a]*Dept. of Computer Science*
*Universidad Carlos III de Madrid, Spain.*
*{david.griol,josemanuel.molina,araceli.sanchis}@uc3m.es*
[b]*Dept. of Languages and Computer Systems,*
*University of Granada, Spain.*
*zoraida@ugr.es*

**Abstract**

In this paper, we present a statistical model for spoken dialog segmentation that decides the current phase of the dialog by means of an automatic classification process. We have applied our proposal to three practical conversational systems acting in different domains. The results of the evaluation show that is possible to attain high accuracy rates in dialog segmentation when using different sources of information to represent the user input. The statistical model developed with human-machine dialog corpora has been applied to human-human conversations and provides a good baseline as well insights in the model limitation.

## 1. Introduction

Speech and natural language technologies allow users to communicate in a flexible and efficient manner, making possible to access applications in which traditional input interfaces cannot be used (e.g. in-car applications, access for disabled persons, etc) [1]. Also speech-based interfaces work seamlessly with small devices (e.g., smartphones and tablets PCs) and allow users to easily interact with robotic agents (e.g., Pepper[1]), invoke local applications or access remote information by means of enhanced devices and advanced conversational interfaces (e.g., Amazon

---

[1]https://www.ald.softbankrobotics.com/en/cool-robots/pepper

Echo[2] and Google Home[3]).

Natural human-computer interaction is a complex problem that requires work on multiple levels, such as automatic speech recognition (ASR), spoken language understanding (SLU), dialog management, and speech synthesis [1, 2]. In recent years, the dialog systems community has focused on developing processing algorithms that can profit from the vast quantities of dialog corpus data that are generated every day. In this field, a machine learning techniques reduce human effort in the knowledge engineering process and development of a new conversational system.

Dialog segmentation can be defined as the process of dividing a dialog by one into (sub-)tasks or phases identifying discourse boundaries using cues (e.g. speaker's intention, topic flow, coherence structure, cohesive devices, etc.). The objective is to detect sequences of turns that accomplish a specific objective inside the dialog flow. These are processed combining different kinds of features provided by the automatic speech recognition and spoken language understanding modules, such as semantic similarities, inter-sentence similarities, entity repetition, word frequency, linguistic features, and prosodic and acoustic characteristics [3].

The identification of the tasks/phases is useful to develop dynamic and user-adapted conversational interfaces. Modeling subdialog structures adapted to each specific task is useful to create extensible interfaces with reusable components that can be plugged to more complex tasks. For example, for excerpts of the dialog that are not strictly related to the application domain (greetings, error recovery, etc.).

The dialog segmentation phase can be performed before dialog management as shown in Figure 1. The dialog manager relies on the estimated dialog task to choose the next system actions.

There is also a wide range of natural language processing applications for which discourse segmentation is relevant. For instance, Angheluta, Busser and Moens adapted a three-step segmentation algorithm for automatic text summarization [4]. Their algorithm uses generic topical cues for detecting the thematic structure of a text using synonymy to reduce the vocabulary.

Different proposals apply discourse segmentation to segment text into different fragments in a preprocessing phase of information retrieval and question-answering systems to improve their operation [5, 6]. Walker applies this kind of techniques for anaphora resolution[7]. Finally, different studies show the benefits of using discourse segmentation for question answering tasks in order to take into account the context for the interpretation and answer questions [8], and also for dialog acts segmentation and classification [9].

---

[2]https://www.amazon.com/echo  
[3]https://madeby.google.com/home/

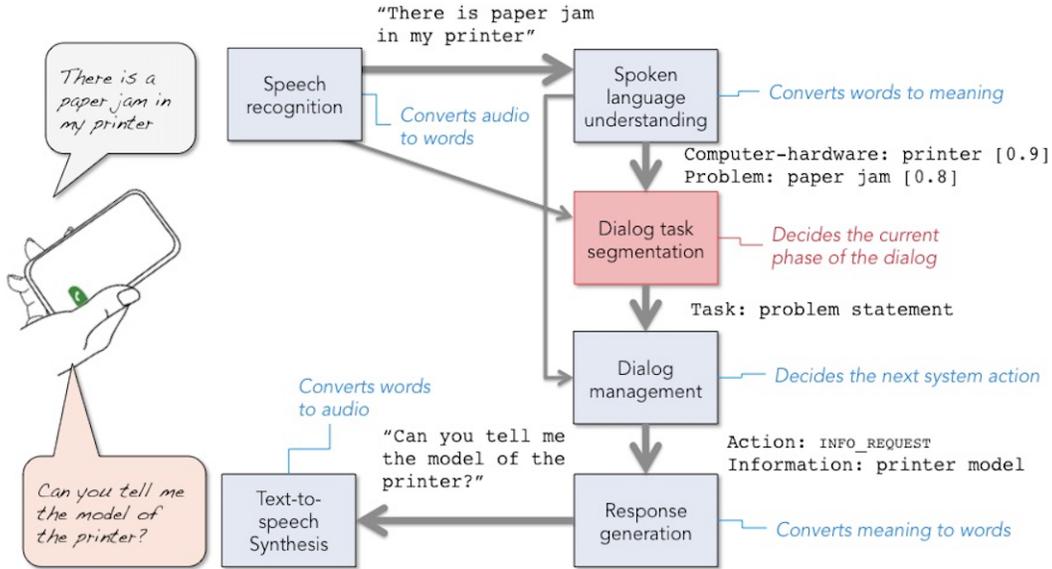

Figure 1: Main modules of a spoken dialog system

In this paper, we describe a machine learning approach for the automatic segmentation of spoken dialogs. Our methodology decides the current phase of the dialog by means of a classification process that considers the complete history of the dialog, which is one of the main advantages regarding the previously described statistical methodologies. Another main characteristic is the inclusion of a data structure that stores the information provided by the user. The main objective of this structure is to easily encode the complete information related to the task provided by the user during the dialog history.

The remainder of the paper is as follows. Section 2 reviews related approaches to dialog structure modeling and segmentation. Section 3 presents our statistical proposal for automatic segmentation of spoken dialogs. Section 4 describes the practical application of this proposal for three different application domains. In Section 5 we discuss the evaluation results. Finally, in Section 6 we present the conclusions and outline guidelines for future work.

## 2. Related work

A number of theories to model discourse structure have been proposed, which are focused on different aspects of dialogs related to the assumptions they made about human-human conversations influenced by many other fields of study including psychology, sociology, philosophy, and computer science.

Most dialog structure models agree that discourse has a compositional structure (i.e., it can be divided into coherent segments that have relationships with each other) [3, 10, 11]. These segments and their relationships constitute the structure of discourse. These two elements can be seen from two different perspectives: an informative perspective and an intentional perspective. The former captures the real content transmitted, that can be modeled by its superficial or semantic representations. The latter captures the speaker's intention behind each statement and the general objective of a discourse.

### 2.1. Informational-oriented discourse structures

The Discourse Representation Theory (DRT) [12] represents the meaning of a text (Discourse Representation Structure, DRS) by means of a logical language similar to the first-order predicate logic, which is extended to represent the context considering the preceding sentences. However, the proposed representation is not focused on the compositional structure of the discourse.

In the Linguistic Discourse Model (LDM) [13], the semantic representation is based on Discourse Parse Trees (DPT) that consider the structural relations and the relations among the contextual categories to capture both propositional content and discourse contexts. Each node in this tree (Discourse Constituent Units, DCU) represents a semantic unit expressing a single event often marked by discourse operators, which denote the relationships among the DCUs. Discourse grammars are usually employed to construct the DPT from DCUs. As in the case of DRT, the LDMs describe semantic representations in terms of truth conditions. In addition, they are influenced by the sentence syntax instead of the objectives of the discourse and the task structure.

The Segmented Discourse Representation Theory (SDRT) [14] extends the DRT model by using a discourse relationship between the current sentence and the previous one. This relationship is used to determine how to combine these two sentences with the analysis of the previous sentences in the overall structure of the semantic representation. A discourse unit is defined at the propositional level and the set of discourse segments recursively defined and their relationships (structural and non-structural) determine how to segment the structure of the discourse. The main difference between this model and the LDM is that the relations in SDRT affect not only the structure of the discourse, but also its semantic representation.

The Rhetorical Structure Theory (RST) [15, 16] explains the structure of a discourse by means of the coherence relationships among its parts. These relationships are classified into subject matter relations (informational) and presentational relations (intentional). This models have been mainly used for text generation and automatic summarization.

DRT, LDM and SDRT differ on how the structure of a discourse affects its semantic representation. They do not explicitly model users' intentions. Although RST considers both the informational and intentional perspectives, the theory also focuses more on the informational perspective and does not consider discourse relations among sentences to create the discourse-level semantic representation.

## 2.2. Intentional-oriented discourse structures

Intentional-oriented theories model the structure of the dialog describing how the user's intentions captured by speech acts or dialog acts fit together in a dialog.

The Speech act theory [17, 18] analyzes the structure of the discourse considering the intention of the user's utterances and the effects on the listener. The set of basic dialog acts can be classified into illocutionary, commissives, expressives and declarations. Many researchers have modified this basic speech act taxonomy to better suite their tasks by adding more domain-specific acts. In addition, dialog acts are usually a key component in other theories modeling the discourse structure, such as dialog grammars, plan-based models and the theory of conversation acts.

The Dialog Act Markup in Several Layers (DAMSL) [19, 20] annotation scheme extends the speech act theory to annotate task-oriented conversations by using multiple labels defined in three orthogonal layers (Forward Communicative Functions, Backward Communicative Functions, and Utterance Features). These layers are respectively used to define a taxonomy similar to the speech act theory, indicate the relationships between the current utterance and the previous ones, and describe the content and form of an utterance. This annotation scheme can be extended with additional domain-specific acts and layers. The main drawback of the model is that the only relationship that is captured is the one between the current utterance and the previous one.

Dialog grammars are based on the detection of the regular patterns that are present in collaborative conversations contains regular patterns, such as question/answer pairs [21]. These hierarchical regular patterns can be expressed using grammars in which the rules specify how the dialog can be segmented into smaller units (e.g., goals, subgoals and dialog acts). Although dialog grammars can be used to predict the next element of a conversation, it is very difficult to generate the complete set of rules that covers all possible variations of conversations in complex domain.

In the Plan-based model [22], dialogs are perceived as a plan that the participants follow to provide appropriate responses and achieve specific goals. The plan describes how the speech acts that model speaker's intentions relate to the conversation goal. The plan-based dialog approach for dialog management has been applied in many complex dialog systems [18]. However, these models usually make strong assumptions about the plan and the environment in which it will be executed (e.g., dialog participants' beliefs do not change), which are not practical in real situations. In addition, plan-based models fail for dialogs that do not follow the task structure closely (e.g. topic changes, clarification or correction subdialogs, etc.). Several augmented plan-based models have been proposed to address these issues.

The theory of conversation acts [23] models the dialog as a set of speaker-hearer actions instead of single agent actions to make grounding actions more explicit. Four levels of actions are defined to maintain the content and coherence of the dialog: turn-taking acts, grounding acts, core speech acts, and argumentation acts. These levels capture distinct dialog information and can be employed independently from each other in a dialog system.

Grosz and Sidner's Theory of discourse structure (GST) [24] models the structure of the dialog using the concepts of discourse unit and discourse coherence. Discourse units, which are defined based on dialog participants' intention, are a set of utterances that fulfills a specific function in the overall goal of the dialog. The intentional structure is used to capture the discourse coherence. Although GST describes an abstract model of discourse structures, the theory does not detail how to solve discourse segmentation, the recognition of discourse segment purposes and the detection of the relationships between them.

The Information State Theory [25] uses the concept of information state to represent the information that can be used to differentiate one dialog from the others. Instead of defining a specific representation for the dialog structure, the theory provides a general framework for dialog modeling that can be implemented in the context of any dialog structure theory.

### 2.3. Data-driven approaches to dialog structure modeling

Research on data-driven approaches to dialog structure modeling is relatively new and focuses mainly on recognizing a structure of a dialog as it progresses [26, 27]. In the literature, there are different methodologies for discourse segmentation and the construction of dialog models including task/subtask information. Unsupervised clustering and segmentation techniques are used in [3] to identify concepts and subtasks in task-oriented dialogs.

Passoneau and Litman [28] presented an algorithm for identifying topic boundaries that uses decision trees to combine multiple linguistic features extracted from corpora of spoken text. These include prosodic features such as pause duration, lexical features such as the presence of certain cue-phrases near boundary candidates, and deeper semantics questions such as whether two noun phrases on opposite sides of a boundary candidate.

Yamron [29] presented an approach to segmentation that models an unbroken text stream as an unlabeled sequence of topics using Hidden Markov Models. Ponte presented in [30] an approach based on information retrieval methods that map a query text into semantically related words and phrases. The approach presented in [31] is based on using adaptive language models and cue-word features for incrementally building an exponential model to extract features that are correlated with the presence of boundaries in labeled training text.

Diverse machine-learning methodologies have been recently proposed for dialog state tracking (DST) [32, 33, 34], a similar task whose objective is to use the system outputs, user's utterances, dialog context and other external information sources to track what has happened in a dialog.

Bayesian dynamic networks are used in generative methods to model a dialog [35]. Unobserved random variables are used to model the true user actions and dialog states, and the probability distribution over dialog states given the system's action and a noisy observation of the true user action is updated using Bayesian inference. The main drawback of these methods are that additional dependencies and structures must be learned to consider potentially useful features of the dialog history. In addition, many independence assumptions must be made to make the models tractable [32].

The parameters for discriminative methods are directly tuned using machine learning and labeled dialog corpus [36]. There are two main categories of discriminative DSTs models: static classifiers and sequence models. The sequence of previous observations is used in static classifiers methods to learn learn a probability distribution for the current state. Different approaches have proposed maximum entropy linear classifiers [36], neural networks [37], and web-style ranking models [38]. Recurrent Neural Networks (RNNs) have been recently proposed as to deal with the high dimensional continuous input features involved in sequential models [39, 40].

## 3. Our proposed methodology for spoken dialog segmentation

The task of detecting the current task of a dialog can be considered as specifying the distribution:

$$P(T_t|o_1, \cdots, o_t)$$

where $T_t$ is the dialog task at the current turn $t$, and $o_1, \cdots, o_t$ is the sequence of observations from the SLU, ASR and machine actions up to and including the current turn.

Spoken dialog systems that help the user to complete a specific task usually define the information that the user and the system can provide in terms of slots. For this reason, they are usually called slot-based dialog systems. The slots and their possible values specify the system's domain, that is, the scope of what users can talk about and the tasks that they can complete. The slots are also related to the set of possible actions that the system can perform, the semantics of the user's utterances and the possible dialog states.

We consider a task-oriented dialog to be the result of incremental creation of a shared plan by the participants [41]. This shared plan consists of several tasks and subtasks. The goal of task segmentation is to predict if the current utterance in the dialog is part of the current task or it starts a new task or subtask. We model this prediction as a maximization problem as the following equation shows:

$$\hat{T}_i = \underset{T_i \in \mathcal{T}}{\operatorname{argmax}} P(T_i | U_1 \cdots U_i, T_1 \cdots T_{i-1}) \quad (1)$$

where set $\mathcal{T}$ contains all the possible tasks/subtasks defined and $U_n$ is the semantic representation of the user utterance at time $n$ in terms of the list of features provided by the Spoken Language Understanding (SLU) module of the conversational agent. The prediction of the current task, that is a local process, takes into account the previous history of the dialog, that is to say, the sequence of user turns and dialog segments preceding time $i$.

The lexical, syntactic and semantic information associated to the speaker's $i$th turn ($U_i$) is usually represented by means of different information sources:

- the words uttered;

- dialog acts, which represent the meaning of an utterance at the level of the speaker's intention in producing that utterance (e.g., *Acceptance*, *Not-Understood*, *Tag-Question*, or *Apology*).

- part of speech tags, also called word classes or lexical categories. Common linguistic categories include noun, adjective, and verb, among others;

- predicate-argument structures, used by SLU modules in various contexts to represent relations within a sentence structure. They are usually represented as triples (subject-verb-object).

- named entities: sequences of words that refer to a unique identifier. This identifier may be a proper name (e.g., organization, person or location names), a time identifier (e.g., dates, time expressions or durations), or quantities and numerical expressions (e.g., monetary values, phone numbers, etc.).

The main problem to resolve Equation 1 is regarding the number of possible sequences of user's utterances preceding the current one, which could be very large in a practical conversational system. To solve this problem, we define a data structure, which we call *User Register* ($UR$), and contains the information provided by the user throughout the previous history of the dialog. The prediction of the current phase of the dialog $T_i$ is then modeled by means of the following equation:

$$\hat{T}_i = \underset{T_i \in \mathcal{T}}{\operatorname{argmax}} P(T_i | UR_i, T_1 \cdots T_{i-1}) \qquad (2)$$

As a practical implementation of this equation, we propose the use of a classification process that takes the semantic information of the user's utterances and the sequence of previous dialog tasks as input, and provides the probability of the dialog being at each of the dialog tasks as output. The set of variables in Equation 2 are codified in the following form:

- The previous dialogs tasks and subtasks ($T_1 \cdots T_{i-1}$): Each dialog task is modeled using a variable, which has as many bits as possible tasks and subtasks defined for each application domain ($C$).

$$\vec{T}_i = (T_{i_1}, T_{i_2}, T_{i_3}, \cdots, T_{i_C}) \in \{0,1\}^C$$

- User register ($UR_i$): As previously stated, the user register includes the information provided by the user until the current moment of the dialog. Each slot related to task-dependent information sources in the $UR$ has been coded in terms of three possible values, $\{0, 1, 2\}$, according to the following criteria: (0) The value of the slot is unknown or it is empty; (1) the value of the slot is known with a confidence score that is higher than a given threshold; (2) the value of the slot has a confidence score that is lower than the given threshold. To decide whether the state of a certain value in the $UR$ is 1 or 2, the dialog system employs confidence measures provided by the ASR and SLU modules [42]. Thus, each one of the task-dependent user dialog acts can take the values $\{0, 1, 2\}$ and then be modeled using a variable with three bits in the same way described for the task-dependent slots.

$$\vec{UR}_i = (UR_{i_1}, UR_{i_2}, UR_{i_3}) \in \{0,1\}^3$$

User's utterances can contain specific values for task-dependent slots, but can also provide other kinds of information, such as task-independent information (for instance, *Affirmation*, *Negation*, and *Not-Understood* dialog acts). These three dialog acts have been coded with the same codification used for the task-dependent information in the $UR$; that is, each one of these three dialog acts can take the values $\{0, 1, 2\}$. This information is modeled using three variables with three bits.

*3.1. Classification functions*

Different classification functions can be defined to solve Equation 2. We have evaluated three different definitions of such a function: a multilayer perceptron (MLP) [43], a decision tree classifier, and a fuzzy-rule-based (FRB) classifier.

- **Decision tree classifier**: Decision trees are widely used as algorithms for problem solving, tools for data mining and knowledge representation, and classifiers that predict the value of the decision attribute for a new object given by values of conditional attributes [44, 45]. The C4.5 decision tree learning algorithm [46] has been used to learn this classification model, using the Weka machine learning software for classifying the complete list of features contained in the user register. The information gain (difference in entropy) is used as splitting criterion to select the attribute to make the decision. Whenever a new attribute is being selected, the algorithm calculates the entropy values for each possible selection, and selects the attribute with the highest information gain. This continues at each node (working top-down) until eventually the classes are split definitively.

- **FRB classifier**: We have recently developed a toolkit to develop dialog managers for spoken dialog systems based on evolving Fuzzy-rule-based (FRB) classifiers [47]. The toolkit uses the *eClass0* (evolving classifier) for the definition of the classification function. These classifiers generate several fuzzy rules per class using an evolving clustering approach to decide when a new rule is created:

$$Rule_i = IF(Feature_1 \ is \ P_1) \ AND \ldots AND (Feature_n \ is \ P_n) THEN \ Class = T_i$$

where $i$ denotes the number of rule; $n$ is the number of input features; the vector $Feature$ stores the observed features, and the vector $P$ stores the values of the features of one of the prototypes of the corresponding class $T_i \in \{\text{set of different tasks/subtasks}\}$.

The *potential* (Cauchy function of the sum of distances between a certain data sample and *all* other data samples in the feature space) is used in the partitioning algorithm. The potential of the $k^{th}$ data sample ($x_k$) is calculated by means of Equation 3 [48].

$$P(x_k) = \frac{1}{1 + \frac{\sum_{i=1}^{k-1} distance(x_k, x_i)}{k-1}} \quad (3)$$

where *distance* represents the distance between two samples in the data space.

The potential can be calculated using the cosine distance to measure the similarity between two samples; as it is described in Equation 4.

$$cosDist(x_k, x_p) = 1 - \frac{\sum_{j=1}^{n} x_{kj} x_{pj}}{\sqrt{\sum_{j=1}^{n} x_{kj}^2 \sum_{j=1}^{n} x_{pj}^2}} \quad (4)$$

where $x_k$ and $x_p$ represent the two samples to measure its distance and $n$ represents the number of different attributes in both samples.

- **MLP classifier**: In the multilayer perceptron classifier, the input layer holds the codification of the input $UR_i, T_1 \cdots T_{i-1}$. The values of the output layer can be seen as an approximation of the a posteriori probability of the input belonging to the associated class representing the current task/subtask.

To train and evaluate the neural networks, we used the *April* toolkit [49]. We firstly tested the influence of the topology of the MLP, by training different MLPs of increasing number of weights using the standard backpropagation algorithm (with a sigmoid activation function and a learning rate equal to 0.2), and selecting the best topology according to the mean square error (MSE) of the validation data. The minimum MSE value was achieved using an MLP of one hidden layer of 32 units. We followed our experimentation with MLPs of this topology, training MLPs with several algorithms (the incremental version of the backpropagation algorithm with and without momentum term, and the quickprop algorithm). The best result on the validation data was obtained using the MLP trained with the standard backpropagation algorithm and a value of LR equal to 0.3.

## 4. Demonstrations with practical dialog systems

As a proof-of-concept we have implemented our proposal in three application domains. We have used three corpora: SoftHard, Dihana and Let's Go!, which main characteristics are summarized in Table 1 and detailed in the following subsections.

Table 1: Main characteristics of the corpora

|  | SoftHard | DIHANA | Let's Go! |
|---|---|---|---|
| Number of dialogs | 150 | 713 | 10,415 |
| Number of user turns | 1,545 | 4,002 | 122,025 |
| Average number of user turns per dialog | 10.3 | 5,6 | 11.7 |
| Number of possible user acts | 42 | 19 | 13 |
| Number of possible system acts | 32 | 29 | 17 |
| Number of main tasks | 14 | 19 | 14 |

Figure 2 shows the scheme follow to test our dialog segmentation technique and the relations of each step with the characteristics of the training corpora explained in the following subsections.

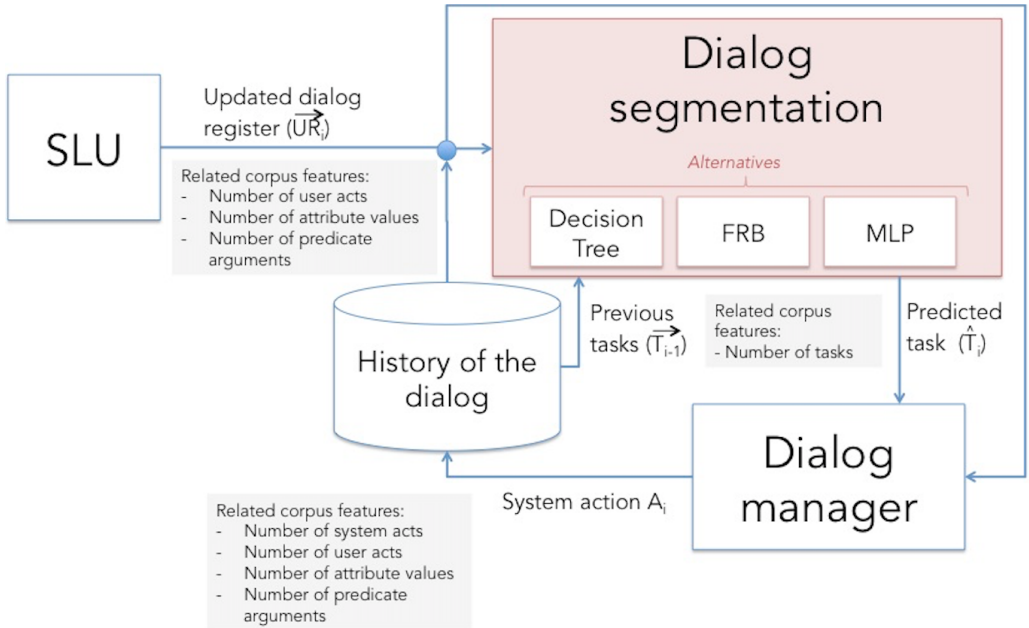

Figure 2: Graphical scheme of the operation of the proposed dialog segmentation technique

*4.1. SoftHard corpus*

The SoftHard corpus is related to a practical spoken dialog system that acts as a customer support service to help solving simple and routine software/hardware repairing problems, both at the domestic and professional levels.

A corpus of 150 conversations with the system attending the calls of users with a software/hardware problem at the City Council of Leganés (Madrid, Spain) was annotated using a multilevel approach similar to the one proposed in the Luna Project [50]. The first levels include segmentation of the corpus in dialog turns, transcription of the speech signal, and syntactic preprocessing with POS-tagging and shallow parsing. The next level consists of the annotation of main information using attribute-value pairs. The other levels of the annotation show contextual aspects of the semantic interpretation. These levels include the predicate structure, the relations between referring expressions, and the annotation of dialog acts.

The attribute-value annotation uses a predefined domain ontology to specify concepts and their relations. The attributes defined for the task include *Concept*, *Computer-Hardware*, *Action*, *Person-Name*, *Location*, *Code*, *TelephoneNumber*, *Problem*, etc.

Dialog act (DA) annotation was performed manually by three annotators on speech transcriptions previously segmented into turns. The DAs defined to label the corpus can be classified into the following categories: i) Core DAs: *Action-request*, *Yes-answer*, *No-answer*, *Answer*, *Offer*, *ReportOnAction*, *Inform*; ii) Conventional DAs: *Greet*, *Quit*, *Apology*, *Thank*; iii) Feedback-Turn management DAs: *ClarificationRequest*, *Ack*, *Filler*; iv) Non interpretable DAs: *Other*.

The original FrameNet[4] description of frame elements was adopted for the predicate-argument structure annotation, introducing new frames and roles related to hardware/software only in case of gaps in the FrameNet ontology. Some of the frames included in this representation are *Telling*, *Greeting*, *Contacting*, *Statement*, *Recording*, *Communication*, *Being operational*, *Change operational state*, etc. Table 2 shows the complete set of features used for the labeling of the different corpora and for the experiments that are presented in this paper.

The basic structure of the dialogs is usually composed by the sequence of the following tasks: *Opening*, *Problem-statement*, *User-identification*, *Problem-clarification*, *Problem-resolution*, and *Closing*. This set of tasks contains a list of subtasks, such as *Problem-description*, *Problem-Request*, *Problem-Confirmation*, *Brand-Identification*, *Model-Identification*, *Help-Request*, *Message-Confirmation*, *Name-Identification*, *Resolution-Confirmation*, etc. The shared plan is represented as a data register that encapsulates the task structure, dialog act structure, attribute-values and predicate-argument structure of utterances. Figure 3 shows an example of the incremental evolution of dialog structure with the complete set of tasks and subtasks. It can be observed the difficulty of correctly detecting the complete structure of the dialog given the number of possible subtasks associated to each main task.

---

[4]https://framenet.icsi.berkeley.edu/fndrupal/

| Dialog Acts | **Core dialog acts** (*Info-request, Action-request Yes-answer, No-answer, Answer, Offer, ReportOnAction, Inform*), **Conventional dialog acts** (*Greet, Quit, Apology, Thank*), **Feedback/Turn management dialog acts** (*ClarificationRequest, Ack, Filler*), **Non interpretable/Non classifiable** (*Other*) |
|---|---|
| **Attribute-value** | Concept, Computer-Hardware, Time-PartoftheDay, Negation, Action, Person-Name, Person-Surname, Location-Institution, Code, Location-Other, Location-TelephoneNumber, Ordinal-Number, Cardinal-Number, Time-RelativeTime, Problem, Person-Position |
| **Predicate information** | Telling, Greeting, Contacting, Statement, Recording, Communication. Being operational, Change operational state, Operational testing, Being in operation. |

Table 2: List of semantic features used for the labeling of the SoftHard corpus

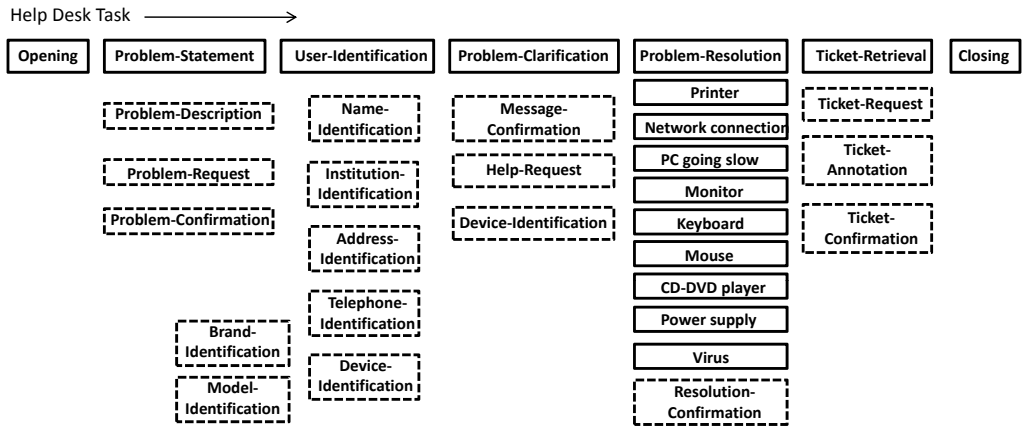

Figure 3: Incremental evolution of the dialog structure

Once finished the *Opening* task (P1), the caller explains the problem the reasons why he/she calls the help-desk during the *Problem-statement* task (P2). In the *User-identification* task (P3), the operator asks for additional information regarding the identity of the caller. Once the caller has described the problem, the operator can ask for additional information to clarify it during the *Problem-clarification* task (P4).

During the *Problem-resolution* task, the operator asks the user to perform specific tests. We have defined nine different subtasks inside this generic segment, given that our goal is to detect not only that the dialog is in this segment, but also what are the specific problem that has to be resolved: *Printer* (P5), *Network connection* (P6), *PC going slow* (P7), *Monitor* (P8), *Keyboard* (P9), *Mouse* (P10), *CD-DVD player* (P11), *Power supply* (P12), and *Virus* (P13). The operator assigns a ticket number for the current call if the problem has not been solved after this task. The user must take note of this number and inform about this to the operator. The dialog ends at the *Closing* task (P14), in which the operator also tries to give a useful advice related to the described problem. Figure 4 shows the transcription of a dialog included in the corpus.

---

**Opening task**
**S1**: Hi. How may I help you?
**Problem-statement task**
**U1**: Hello, my printer does not print correctly.
**User-identification**
**S2**: Tell me your name and surname.
**U2**: My name is Ricardo Santamaria.
**S3**: Can you tell me your phone extension?
**U3**: Nine four sixty-three.
**Problem-clarification task**
**S4**: Can you tell me the brand and model?
**U4**: HP Officejet Pro 8610.
**Problem-resolution task: Printer problem**
**S5**: Is the printer on now?
**U5**: It is on.
**S6**: Can you tell me the message on the printer display?
**U6**: Paper jam is written.
**S7**: Can you see the active printer from the control panel of your PC?
**U7**: It is active.
**S8**: Please try the following steps. Say next after each step. Turn the printer off, and then on again.
**U8**: Next.
**S9**: Print a Printer Status Report.
**U9**: Next.
**S10**: Has this process solved your problem?
**U10**: No.
**S11**: Remove any loose paper from the main and optional trays.
**U11**: Next.
**S12**: Clear any jammed paper from the rear of the printer.
**U12**: Next.
**S13**: Has this process solved your problem?
**U13**: Yes.
**Closing task**
**S14**: Anything else?
**U14**: No, thank you.
**S15**: Most of these steps will also work for troubleshooting an office copier. Have a nice day!

---

Figure 4: Example of a Human-Machine dialog (translation from Spanish to English)

*4.2. DIHANA corpus*

A set of 713 dialogs was acquired in the DIHANA project, whose goal was the development of a dialog system providing railway information using spontaneous speech in Spanish [51]. Although this corpus was acquired using a Wizard of Oz technique (WOz), real speech recognition and understanding modules were used.

The dialogs in the corpus were labeled in terms of dialog acts. In the case of user turns, the dialog acts correspond to the use of frames to represent the meaning of the utterance. For the DIHANA task, eight concepts and ten attributes were defined. The eight concepts are divided into two groups:

1. *Task-dependent concepts*: they represent the concepts the user can ask for (*Hour*, *Price*, *Train-Type*, *Trip-Time*, and *Services*).
2. *Task-independent concepts*: they represent typical interactions in a dialog (*Affirmation*, *Negation*, and *Not-Understood*).

Three levels were defined for the labeling of the 29 system responses. The first level describes the general acts of any dialog, independently of the task. The second level represents the concepts and attributes involved. The third level represents the values of the attributes given in the turn. The following labels were defined for the first level: *Opening*, *Closing*, *Undefined*, *Not-Understood*, *Waiting*, *New-Query*, *Acceptance*, *Rejection*, *Question*, *Confirmation*, and *Answer*. The labels defined for the second and third level were the following: *Departure-Hour*, *Arrival-Hour*, *Price*, *Train-Type*, *Origin*, *Destination*, *Date*, *Order-Number*, *Number-Trains*, *Services*, *Class*, *Trip-Type*, *Trip-Time*, and *Nil*.

Figure 5 shows an example of the semantic interpretation of a user's utterance and a system response.

The dialogs in the corpus were divided into 19 main tasks, which are summarized in Table 3.

*4.3. Let's Go corpus*

Let's Go is a spoken dialog system developed by the Carnegie Mellon University to automatically provide bus schedule information in Pittsburgh for a subset of 5 routes and 559 bus stops. The system has had many users since it was made available for the general public in 2005, with more than 20,000 calls collected just from March to December of 2005 [52], so there is a substantial dataset that can be used to train a dialog model. In addition, this large amount of data from spoken interactions has been acquired with real callers, rather than lab testers. A corpus of 10,415 dialogs with 122,025 dialog acts was distributed among the scientific community for the Dialog State Tracking Challenge (DSTC)[5].

---

[5]https://www.microsoft.com/en-us/research/event/dialog-state-tracking-challenge/

**Output sentence:**
[SPANISH] Sí, me gustaría conocer los horarios para esta tarde desde Valencia.
[ENGLISH] *Yes, I would like to know the timetables for this evening leaving from Valencia.*
**Semantic interpretation:**
(*Affirmation*)
(*Hour*)
   *Origin*: Valencia
   *Departure-Date*: Today
   *Departure-Hour*: Evening

**Output sentence:**
[SPANISH] ¿Quiere conocer los precios saliendo desde Valencia?
[ENGLISH] *Do you want to know the price leaving from Valencia?*
**Semantic interpretation:**
(*Confirmation*:*Price*:*Nil*)(*Confirmation*:*Origin*:*Origin*)

Figure 5: An example of the labeling of a user and system turns in the DIHANA corpus

Table 3: Main tasks for the Dihana dialog system

| | |
|---|---|
| *Welcome*, | *Ask_for_query*, |
| *Confirm_query*, | *Timetables_query_ask_for_attribute*, |
| *Timetables_query_confirm_attribute*, | *Timetables_query_provide_results*, |
| *Fares_query_ask_for_attribute*, | *Fares_query_confirm_attribute*, |
| *Fares_query_provide_results*, | *TriptTime_query_ask_for_attribute*, |
| *TripTime_query_confirm_attribute*, | *TripTime_query_provide_results*, |
| *TypesTrain_query_ask_for_attribute*, | *TypesTrain_query_confirm_attribute*, |
| *TypesTrain_query_provide_results*, | *Services_query_ask_for_attribute*, |
| *Services_query_confirm_attribute*, | *Services_query_provide_results*, |
| *Goodbye* | |

Figure 6 shows an example of a dialog extracted from the Let's Go corpus [53]. Each call to the system starts with a welcome message that prompts the user to make a request. Then, the system waits for the user's response and grabs concepts such as question type (e.g., *When is the next bus to X?*, *How can I go from X to Y?*) or departure and arrival times and places. To be successful, calls require three or four pieces of information from the user: a departure stop, a destination, a travel time, and, optionally a bus route. Stops can be specified in one of three ways:

> **System**: Welcome to the Pittsburgh bus information line. You can say start over anytime. Where are you going to take your bus from?
> **User**: Hi yeah I want to go from McKeesport
> **System**: Sure. Where do you want to go?
> **User**: Duquesne.
> **System**: Did you mean Duquesne?
> **User**: Yes
> **System**: At what time do you want to take it? Say next bus or a time for example 8.40pm.
> **User**: Around 4.15pm
> **System**: I'm going to search for buses around 4.15pm. Is that OK?
> **User**: Yes.
> **System**: On which day?
> **User**: Today
> **System**: I have found one bus. It is at 4.15pm. There is a bus number 61C from McKeesport. It will arrive in Duquesne at 4.21pm.
> **System**: Is there anything else I can do for you?
> **User**: No thanks
> **System**: Thanks for calling. Good bye!

Figure 6: Example of dialog extracted from the Let's Go corpus

the nearest intersection to the stop (e.g., Forbes [Avenue] at Murray [Avenue]), a neighborhood (e.g., Oakland), or a landmark or other point of interest (Pittsburgh International Airport, Waterworks Mall). The system explicitly prompts the user to provide the missing information to complete the query. Once the system has the required information to answer the user's query, it submits a query to the database, presents the results to the user, and prompts for a new query.

With regard the semantic representation defined for the task, the system uses a set of user dialog acts that has been classified into 16 categories following the criteria described in [54]. A total of 16 categories of user dialog acts were defined. Four of the dialog acts are used to model where the user is leaving from (monument, pair of road names, neighborhood, or stop). The four dialog acts used for modeling the place of arrival are similar. Six dialog acts are used for describing the user's required time of travel (next bus or specific times). The *meth* node describes whether the user is asking for a bus with some constraints, is finished or wants to restart. The dialog act *disc* models how the user issues "discourse" actions, which relate to only one turn in a dialog.

A total of 36 system dialog acts were defined. These dialog acts can also be classified into 5 groups: *formal* (dialog formalities like "welcome"), *results* (presentation of search results), *queries* (request for values to fill slots), *statusreports* (when the system reports about its status, e.g. "looking up database"), *error* (error messages), and *instructions* (instructions to the user how to speak to the system).

The different objectives of the dialogs for the Spoken Dialog Challenge were labeled in the corpus by considering the different places and times for which the users required information (from one to five), users' requirements about previous and next buses, number of uncovered places, and possible system failures. The different combinations of these parameters in the corpus lead to the definition of 38 different objectives. The dialogs were divided into 14 main tasks, which are shown in Table 4.

Table 4: Main tasks for the Let's Go! dialog system

| | |
|---|---|
| *Welcome*, | *Ask_for_query*, |
| *Ask_for_query*, | *Ask_for_help*, |
| *Confirm_query*, | *Confirm_attribute*, |
| *Looking_up_database*, | *Restart_dialog*, |
| *New_query*, | *Provide_results*, |
| *Error*, | *Provide_instructions*, |
| *Query_error*, | *Goodbye* |

## 5. Results and discussion

The developed methodology for the task detection has been evaluated by means of the dialogs of the corpora described in the previous section. For the three corpora, the evaluation of the statistical dialog segmentation technique was carried out *turn by turn* using a five-fold cross validation process. Each one of the two corpus was randomly split into five subsets. Each trial used a different subset taken from the five subsets as the test set, and the remaining 80% of the dialogs was used as the training set. The three proposals for the definition of the classification function described in Section 3.1 have been evaluated. Tables 5, 6 and 7 compile the results (precision, recall and F-Measure) obtained in the recognition of the tasks for each corpus. The tables show the average results for the main tasks of each system considering all their subtasks.

Table 5: Task and subtasks prediction for the SoftHard dialog system

| | Decision tree classifier | | | FRB-classifier | | | MLP-classifier | | |
|---|---|---|---|---|---|---|---|---|---|
| | Prec. | Rec. | F-Meas. | Prec. | Rec. | F-Meas. | Prec. | Rec. | F-Meas. |
| Opening | 0.98 | 0.95 | 0.97 | 0.98 | 0.93 | 0.95 | 0.99 | 0.94 | 0.96 |
| Prob-statement | 0.94 | 0.90 | 0.92 | 0.95 | 0.91 | 0.93 | 0.97 | 0.92 | 0.94 |
| User-identification | 0.93 | 0.89 | 0.91 | 0.94 | 0.89 | 0.91 | 0.96 | 0.92 | 0.94 |
| Prob-clarification | 0.91 | 0.88 | 0.90 | 0.92 | 0.88 | 0.90 | 0.95 | 0.91 | 0.93 |
| Prob-resolution | 0.90 | 0.86 | 0.88 | 0.91 | 0.87 | 0.89 | 0.94 | 0.90 | 0.93 |
| Closing | 0.96 | 0.94 | 0.95 | 0.97 | 0.92 | 0.94 | 0.98 | 0.96 | 0.97 |

Table 6: Task and subtasks prediction for the DIHANA dialog system

|                  | Decision tree classifier |      |         | FRB-classifier |      |         | MLP-classifier |      |         |
|------------------|-------|------|---------|-------|------|---------|-------|------|---------|
|                  | Prec. | Rec. | F-Meas. | Prec. | Rec. | F-Meas. | Prec. | Rec. | F-Meas. |
| Welcome          | 0.97  | 0.89 | 0.93    | 0.98  | 0.90 | 0.94    | 0.98  | 0.94 | 0.96    |
| Ask_for_query    | 0.88  | 0.82 | 0.85    | 0.89  | 0.83 | 0.86    | 0.91  | 0.88 | 0.90    |
| Confirm_query    | 0.83  | 0.78 | 0.80    | 0.85  | 0.79 | 0.82    | 0.86  | 0.84 | 0.85    |
| Timetable_query  | 0.81  | 0.77 | 0.79    | 0.81  | 0.77 | 0.79    | 0.82  | 0.78 | 0.80    |
| Fares_query      | 0.80  | 0.74 | 0.77    | 0.81  | 0.75 | 0.78    | 0.82  | 0.77 | 0.79    |
| TripTime_query   | 0.81  | 0.74 | 0.77    | 0.81  | 0.74 | 0.77    | 0.82  | 0.78 | 0.80    |
| TypesTrain_query | 0.83  | 0.72 | 0.77    | 0.83  | 0.72 | 0.77    | 0.84  | 0.80 | 0.82    |
| Services_query   | 0.82  | 0.71 | 0.76    | 0.82  | 0.71 | 0.76    | 0.84  | 0.79 | 0.81    |
| Provide_results  | 0.87  | 0.81 | 0.84    | 0.89  | 0.83 | 0.86    | 0.91  | 0.87 | 0.89    |
| Goodbye          | 0.94  | 0.90 | 0.92    | 0.96  | 0.92 | 0.94    | 0.97  | 0.95 | 0.96    |

Table 7: Task and subtasks prediction for the Let's Go! dialog system

|                      | Decision tree classifier |      |         | FRB-classifier |      |         | MLP-classifier |      |         |
|----------------------|-------|------|---------|-------|------|---------|-------|------|---------|
|                      | Prec. | Rec. | F-Meas. | Prec. | Rec. | F-Meas. | Prec. | Rec. | F-Meas. |
| Welcome              | 0.98  | 0.93 | 0.95    | 0.98  | 0.94 | 0.96    | 0.98  | 0.97 | 0.98    |
| Ask_for_query        | 0.88  | 0.82 | 0.85    | 0.85  | 0.82 | 0.84    | 0.92  | 0.90 | 0.91    |
| Ask_for_attribute    | 0.87  | 0.83 | 0.85    | 0.88  | 0.84 | 0.86    | 0.90  | 0.88 | 0.89    |
| Confirm_query        | 0.88  | 0.82 | 0.85    | 0.89  | 0.84 | 0.86    | 0.91  | 0.87 | 0.89    |
| Confirm_attribute    | 0.87  | 0.82 | 0.84    | 0.88  | 0.84 | 0.86    | 0.90  | 0.86 | 0.88    |
| Provide_results      | 0.92  | 0.86 | 0.89    | 0.93  | 0.88 | 0.90    | 0.95  | 0.92 | 0.94    |
| Provide_instructions | 0.82  | 0.77 | 0.79    | 0.83  | 0.79 | 0.81    | 0.86  | 0.81 | 0.83    |
| Query_error          | 0.84  | 0.79 | 0.81    | 0.85  | 0.80 | 0.82    | 0.87  | 0.84 | 0.86    |
| Goodbye              | 0.96  | 0.92 | 0.94    | 0.96  | 0.94 | 0.95    | 0.98  | 0.96 | 0.97    |

As can be observed in the tables, the best results for all corpora are obtained with the MLP classifier. Table 8 presents the average precision, recall and F-measure obtained for all tasks in each corpus using the MLP classifier.

|                  | Avg. Precision | Avg. Recall | Avg. F-measure |
|------------------|----------------|-------------|----------------|
| SoftHard system  | 0.97           | 0.93        | 0.95           |
| Dihana system    | 0.88           | 0.84        | 0.86           |
| Let's Go! system | 0.92           | 0.89        | 0.90           |

Table 8: Comparison of results of the approach to dialog segmentation for the three corpora

The table shows that, despite of having a higher number of tasks, the homogenous structure of the SoftHard dialogs makes tasks more predictable, and so the accuracy for the segmentation is higher. The main problem for Dihana is the lower number of samples for some tasks (*triptimes*, *typesoftrains*, and *services*), this makes the dialogs containing such tasks more difficult to segment and has lowered the average accuracy rates. With respect to the Let's Go! system, the main difficulty detected occurred in the occasions when users wanted to restart the dialog, asked for help or there was an error in the communication. These tasks are very difficult to detect as they are less predictable. Albeit these challenges, the accuracy and precision obtained for all tasks is high.

5.1. Influence of the use of human-human vs. human-machine dialogs and training features

An additional corpus of 150 human-human dialogs was available for the SoftHard system. Figure 7 shows the distribution of dialog segments annotated in each corpus. As can be seen, the tasks distribution is very different to the human-machine corpus described in Section 4.1, and human-human dialogs present an additional 27.31% percentage of situations that has been labeled as *Out of the Task* (P15).

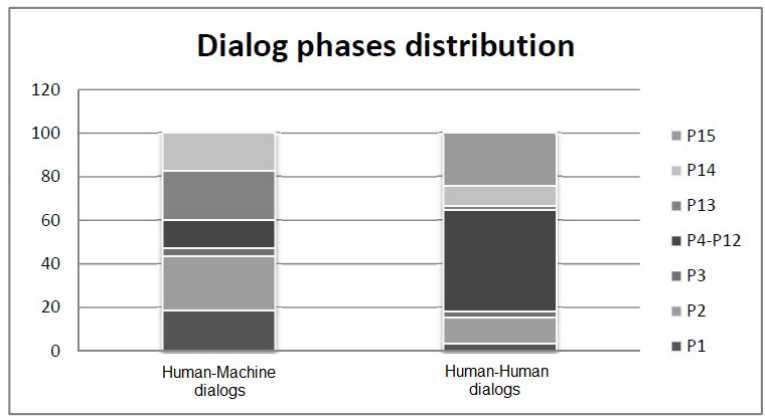

Figure 7: Distribution of dialog segments annotated in the human-machine and human-human dialogs for the SoftHard system

As human-human dialogs are spontaneous, they present several differences with respect to human-machine dialogs. The main one is the great difference in the average number of turns (16.18 turns in the human-machine corpus and 25.71 for the human-human dialogs). This is because human-human dialogs present other minor topics (like small talks about other persons, previous problems, holidays, etc), a high frequency of interruptions, cut-off phrases, and overlapped contributions. This makes that the 18.24% of the utterances of the human-human corpus have been labeled as *Out of the Task*.

Analyzing the annotation available for the DA level, we measured that in average an HH dialog is composed of 37.9±7.3 (Std. Dev.) DAs, whereas a human-machine dialog is composed of 21.9±5.4. The difference between average lengths shows how human-human spontaneous speech can be redundant, while human-machine dialogs are more limited to an exchange of essential information. The standard deviation of a conversation in terms of DAs is considerably higher in the human-human corpus than in the human-machine ones. This can be explained by the fact that the human-machine dialogs follow a unique, previously defined task-solving strategy that does not allow digressions.

The evaluation of the statistical dialog segmentation technique was also carried out *turn by turn* using a five-fold cross validation process. Table 9 shows the results, which indicate that the prediction is improved once the different SLU features are incorporated to the model. As can be seen, the proposed methodology successfully adapts to the requirements of the human-machine dialogs, since a 0.95 F-measure is obtained, measuring the dialog segments provided by the developed module that are equal to the segment annotated in the corpus for these dialogs. This value is reduced to 0.79 for the human-human dialogs, since the *Out of the Task* class is usually confused with the rest of dialog segments related to the task. Therefore, the methodology adapts to the very different nature that has been described for both kind of dialogs.

Finally, we learned a model with the total 150 human-machine dialogs and evaluated it using the total 150 human-human dialogs. This experimentation was designed to evaluate if a model learned with human-machine dialogs can detect the task-related structure of spontaneous human-human conversations. The main challenge of this experiment is that only a maximum of 81.76% can be achieved due to the 18.24% *Out of the task* that is only present in the human-human corpus. As can be observed, the model successfully adapts to detect the task-related parts in the human-human dialogs, achieving a 0.67 F-measure.

|  | Precision | Recall | F-measure |
|---|---|---|---|
| **Human-machine corpus for learning and evaluating** | | | |
| Attribute-Values | 0.89 | 0.87 | 0.88 |
| DAs + Attribute-Values | 0.94 | 0.92 | 0.93 |
| Complete set | 0.97 | 0.93 | 0.95 |
| **Human-human corpus for learning and evaluating** | | | |
| Attribute-Values | 0.72 | 0.60 | 0.66 |
| DAs + Attribute-Values | 0.86 | 0.71 | 0.78 |
| Complete set | 0.87 | 0.72 | 0.79 |
| **Human-machine corpus for learning and human-human corpus for evaluating** | | | |
| Attribute-Values | 0.62 | 0.55 | 0.58 |
| DAs + Attribute-Values | 0.69 | 0.61 | 0.65 |
| Complete set | 0.73 | 0.63 | 0.67 |

Table 9: Average results of the evaluation of the proposed dialog segmentation technique for the help desk system

## 6. Conclusions and future work

In this paper, we have presented a statistical approach for automatic dialog segmentation in conversational interfaces. This approach uses feature selection to collect a set of informative features into a model that includes the information provided by the user during the complete history of the dialog. This model can be used to predict the current task in the dialog, helping the dialog manager in the selection of the next system prompt. The experimental results show that the statistical approach successfully adapts to the requirements of three different applications domains, obtaining high accuracy and precision rates, not only separately for human-machine and human-human dialogs acquired for this task, but also when training the system with human-machine dialogs and testing it with spontaneous human-human dialogs. For future work, we want to perform a more detailed analysis of the situations that have been labeled as *Out of the Task*, studying if our proposal is able to differentiate them. We also plan to incorporate additional information regarding the user, such as specific user profiles adapted to each specific interaction domain. Finally, we also intend to analyze the impact of the corpus size in the results obtained.